# Overview of Linacs

*A.M. Lombardi*
CERN, Geneva, Switzerland

**Abstract**
In this paper, we give an overview of the different types of linac accelerators, with special emphasis on their use for a hadron-therapy facility.

**Keywords**
Linacs; medical.

## 1 Introduction

Linac stands for LINear ACcelerator: a single pass device that increases the energy of a charged particle by means of an (radio frequency, RF) electric field and is equipped with magnetic elements (quadrupoles, solenoids, bending magnets) to keep the charged particle on a given trajectory. The motion equation of a charged particle in an electromagnetic field can be written as

$$\frac{d\vec{p}}{dt} = q \cdot \left( \vec{E} + \vec{v} \times \vec{B} \right), \tag{1}$$

where

$\vec{p} = $ momentum $= \gamma m_0 \vec{v}$,

$q, m_0 = $ charge, mass,

$\vec{E}, \vec{B} = $ electric field, magnetic field,

$t = $ time,

$\vec{x} = $ position vector, and

$\vec{v} = \dfrac{d\vec{x}}{dt} = $ velocity.

If we rewrite Eq. (1) in a slightly different form, we can identify the key parameters of a linear accelerator. The relativistic factor gamma on the left-hand side of Eq. (2) indicates whether we are in the non-relativistic, semi-relativistic, or fully relativistic regime. The factor $q/m_0$ on the right-hand side indicates what type of particle the linac is designed for, and the $E$ term and $B$ term on the right-hand side indicate the type of RF structure and the type of focusing, respectively.

$$\frac{d}{dt}\left( \gamma \frac{d\vec{x}}{dt} \right) = \frac{q}{m_0} \cdot \left( \vec{E} + \frac{d\vec{x}}{dt} \times \vec{B} \right). \tag{2}$$

It is interesting to recall the dependence of the relativistic beta for electrons and protons, both particles being used in medical linacs. From Fig.1, we can see that the electrons are relativistic at a few MeV of energy, whereas the protons (and the carbon ions) are never fully relativistic in the energy range interesting for hadron therapy (250 MeV for protons and 450 MeV/u for carbon ions).

Electron linacs for medical applications are very compact. They operate in the energy range between 4 and 25 MeV and they are, nowadays, commercially available. They are, therefore, not the subject of this lecture.

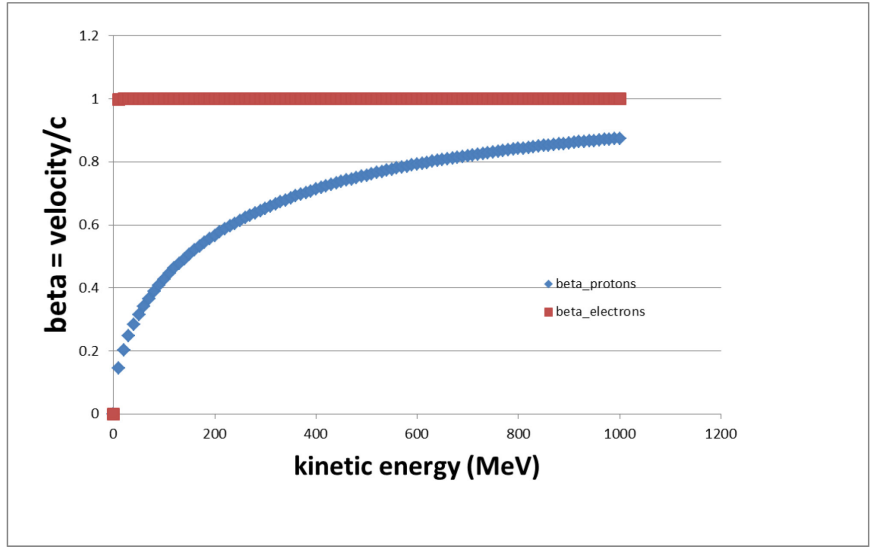

| Proton energy (MeV) | Relativistic beta | Relativistic gamma |
|---|---|---|
| 7 | 0.12 | 1.01 |
| 250 | 0.6 | 1.26 |
| 450 | 0.74 | 1.48 |

**Fig. 1:** Relativistic beta for electrons (red) and protons (blue)

The predecessors of modern RF hadron linacs are electrostatic linacs, based on a static field to accelerate particles. This type of linac, used until not long ago in the first stage of acceleration before the dimensions of the system became impractical, is limited to energies of a few MeV. Acceleration by a time-varying electromagnetic field overcomes the limitation of static acceleration. The first experiment towards an RF linac was done by the Norwegian physicist, R. Wideroe [1], in 1928 based on a proposal by Ising dated 1925 [2]. A group {please leave bunch]of potassium ions was accelerated to 50 keV in a system of drift tubes in an evacuated glass cylinder. The available generator provided 25 keV at 1 MHz. It was not until 1931 that the first linac was developed by Sloan and Lawrence at Berkeley Laboratory [3]. The man who brought the linac from an experiment to a practical accelerator was Luis Alvarez. He realized that to proceed to higher energies it was necessary to increase, by an order of magnitude, the frequency and to enclose the drift tubes in an RF cavity (resonator). A 200 MHz 12 m-long Drift Tube Linac (DTL), built by Luis Alvarez at the University of California in 1955 [4], accelerated protons from 4 to 32 MeV. The development of the first linac was made possible by the availability of high-frequency power generators developed for radar application during World War II.

## 2   Linacs for medical applications

In this section, we will explore in more detail hadron linacs for medical applications. The parameter $\gamma$ of Eq. (2) is in the range 1 to 1.48 and the $q/m$ goes from 1 (protons) to 1/3 for carbon ions. In this framework, the particles are non-relativistic (or semi-relativistic towards the end of the acceleration), therefore, the choice of the type of RF structure is critical for the efficiency (and cost) of the accelerator complex. In Table 1, the different types of RF structures used in medical facilities (existing and in the design stage) are listed.

Table 1: RF structures used in existing and planned facilities

| Type of structure | Used in |
|---|---|
| Radio Frequency Quadrupole | HIT,CNAO, MEDAUSTRON, ADAM,TULIP2.0 |
| Interdigital-H structure | HIT CNAO MEDAUSTRON |
| Drift Tube Linac | IMPLART (ENEA FRASCATI) / TULIP2.0 |
| Cell Coupled Linac also called Side Coupled Linac | ADAM, TULIP |

## 2.1 Transverse electric or transverse magnetic modes, and cavity modes

Let us recall the Maxwell equation for $E$ and $B$ fields [5]:

$$\left(\frac{\partial^2}{\partial x^2}+\frac{\partial^2}{\partial y^2}+\frac{\partial^2}{\partial z^2}-\frac{1}{c^2}\frac{\partial^2}{\partial t^2}\right)\vec{E}=0. \qquad (3)$$

In free space, the solution of the above equation shows that electromagnetic fields are of the transverse electromagnetic, TEM, type: the electric and magnetic field vectors are perpendicular to each other and to the direction of propagation. In a bounded medium, e.g. a cavity, the solution of the equation must satisfy the boundary conditions:

$$\vec{E}_{//}=\vec{0} \text{ and } \vec{B}_{\perp}=\vec{0}; \qquad (4)$$

therefore, only either transverse electric (TE) or transverse magnetic (TM) modes are possible.

In TE modes, the electric field is perpendicular to the direction of propagation, whereas in TM modes, the magnetic field is perpendicular to the direction of propagation. In a cylindrical cavity, they are denoted as $TE_{nml}$ and $TM_{nml}$, respectively, where the indices $n$, $m$, and $l$ refer to the azimuthal, radial, and longitudinal components.

Figure 2 shows the first two TE modes in a cylindrical cavity. The cut is perpendicular to the direction of propagation of a beam, the lines represent the electric field, and the dots and the crosses are the points where the magnetic field, parallel to the direction of propagation, enters/exits from the paper.

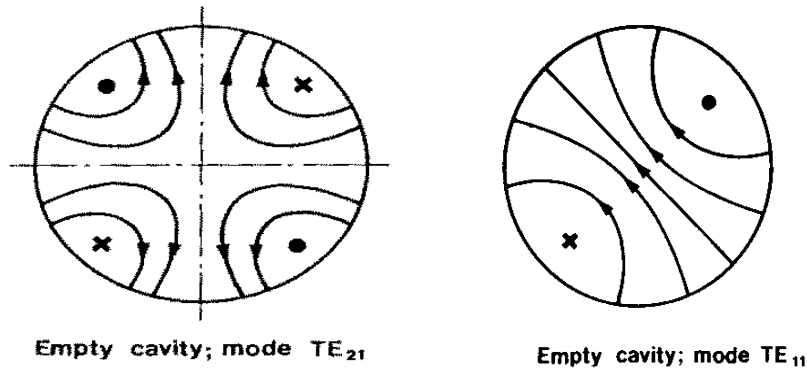

**Fig. 2:** The first two TE modes in a cylindrical cavity: dipole mode $TE_{110}$ and quadrupole mode $TE_{210}$

Figure 3 shows the first TM mode in a cylindrical cavity. On the left-hand side, a perpendicular cut where the beam enters the cavity from left to right, and on the right-hand side, a transverse cut where the beam enters into the page.

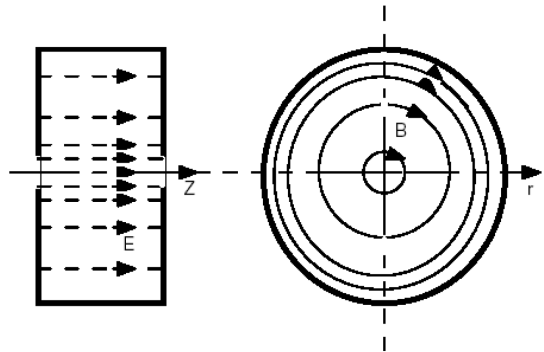

**Fig. 3:** The $TM_{010}$ mode: most commonly used accelerating mode

A linac is composed of a string of cavities, depending on the phase shift of the field in one cavity with respect to the adjacent cavity, we can identify three main cavity modes. These are:

– 0-mode: zero-degree phase shift from cell to cell, so fields in adjacent cells are in phase. The best example is a DTL;
– $\pi$-mode: 180-degree phase shift from cell to cell, so fields in adjacent cells are out of phase. The best example is multi-cell superconducting cavities;
– $\pi/2$ mode: 90-degree phase shift from cell to cell. In practice, these are bi-periodic structures with two kinds of cells, accelerating cavities, and coupling cavities. The CCL operates in a $\pi/2$ structure mode. This is the preferred mode for very long multi-cell cavities, because of very good field stability.

## 2.2 The radio frequency quadrupole

A radio frequency quadrupole (RFQ) [6, 7, 8] is composed of a cavity loaded with four electrodes and therefore, forced to resonate in the $TE_{210}$ mode. A sketch is shown in Fig. 4.

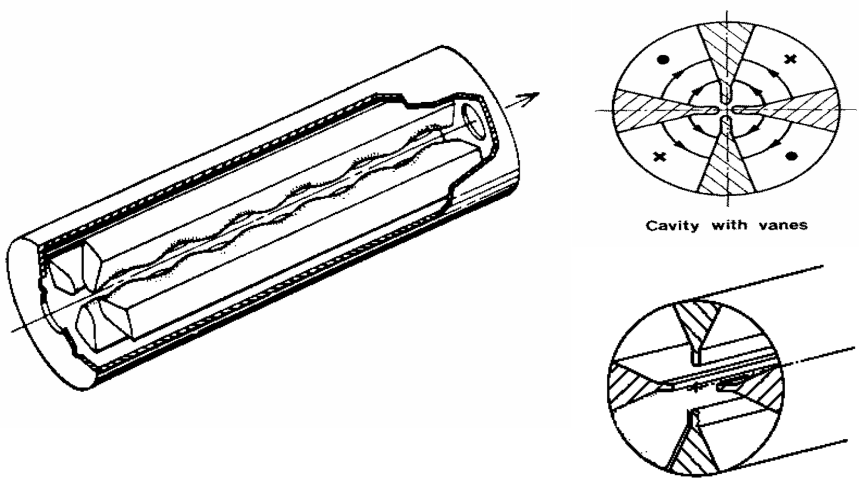

**Fig. 4:** Sketch of an RFQ cavity, courtesy of T. Wangler

The transverse field in an RFQ provides an alternating-gradient focusing structure with a period length equal to $\beta\lambda$, where $\beta$ is the velocity of the particle and $\lambda$ is the RF wavelength. This can be more easily understood by looking at the sketch in Fig.5 and 6. Let us assume that the top and bottom electrodes have positive polarity during the time DT1, DT3…, whereas the left and right electrodes have negative polarity during this time. A positively charged ion beam will pushed away from the top/bottom electrodes and pulled towards the left/right. During this time (DT1, DT3...), the RFQ behaves like a defocusing quadrupole. Conversely, during the time intervals DT2, DT4…, the RFQ behaves like a focusing quadrupole. No force is exerted on the beam at the zero crossing between positive and negative electric fields. During any of the time intervals DT, the beam has travelled a distance equal to $\beta\lambda/2$ and therefore has seen a smooth focusing channel with period $\beta\lambda$.

Acceleration in an RFQ is achieved by periodically modulating the electrodes in the longitudinal direction, as shown in Fig. 7. The periodic longitudinal modulation, which is 180 degrees out of phase in the top/bottom electrodes with respect to the left/right pair, deforms the pattern of the pure TE mode by creating a longitudinal component proportional to the depth of the modulation. The synchronism between the increasing velocity of the particle and the longitudinal component can be controlled with the wavelength of the modulation.

The RFQ is the accelerator that has bridged the gap between a proton/hadron source and a conventional (TM mode) accelerator. Its strong points are the electric focusing which allows a low-energy beam to be accepted, and the adiabatic bunching which preserves beam quality and allows high capture. These two features combined have increased the efficiency of the very first phase of pre-acceleration from 50% to more than 90%, also in the presence of a strong space charge. Besides, as the transverse and longitudinal dynamics are machined in the electrode microstructure, the RFQ is very easy to operate in routine runs of an accelerator complex.

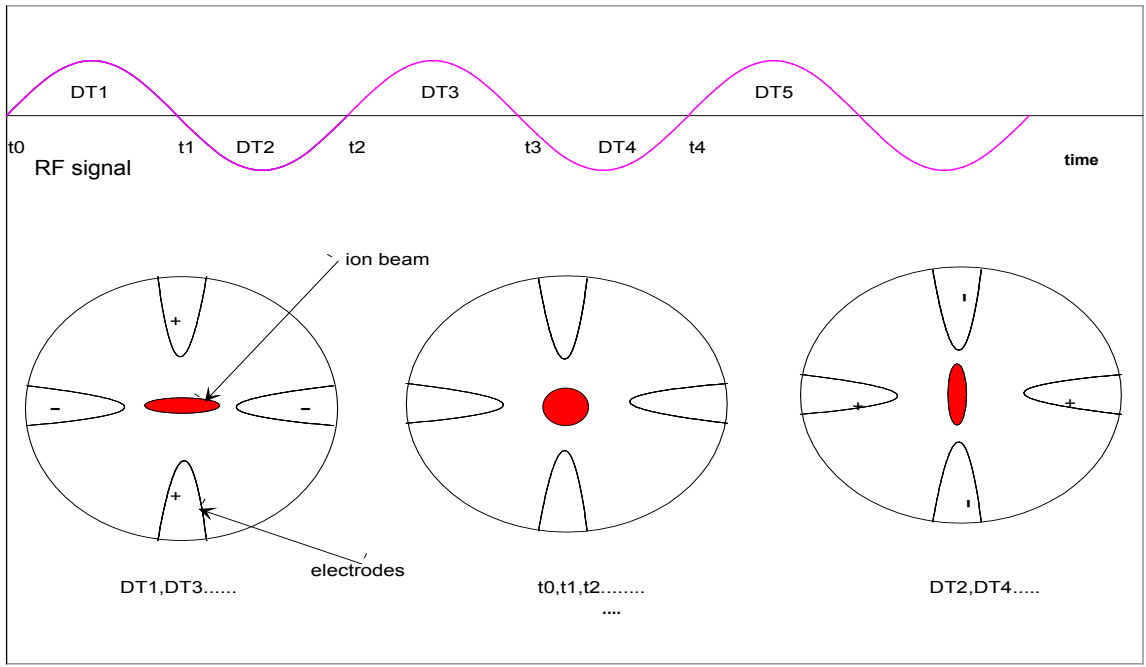

**Fig. 5:** Time-varying transverse focusing field in an RFQ

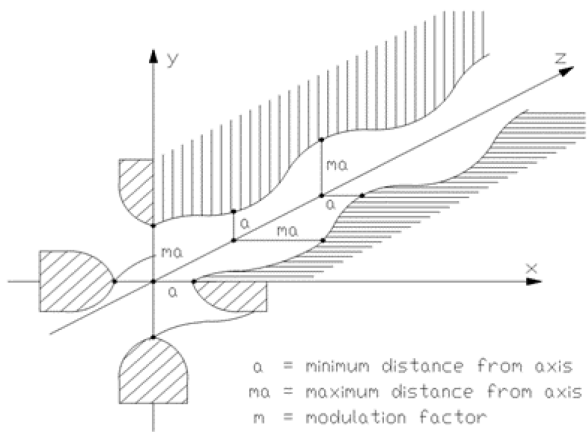
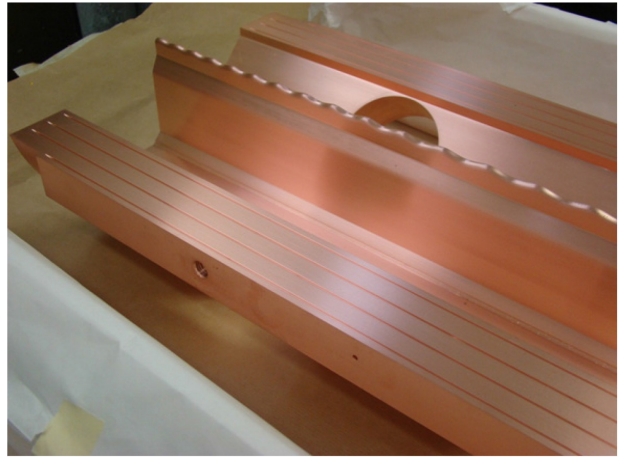

**Fig. 6:** Longitudinal modulation on the RFQ electrodes

## 2.3 Interdigital H structure

A structure used in all existing hadron-therapy facilities is the Interdigital H (IH) structure. A natural structure to follow the RFQ in low-current machines, it is composed of drift tubes alternately held by lateral stems, and sections including magnetic quadrupoles for the transverse focusing, as shown in Fig. 7.

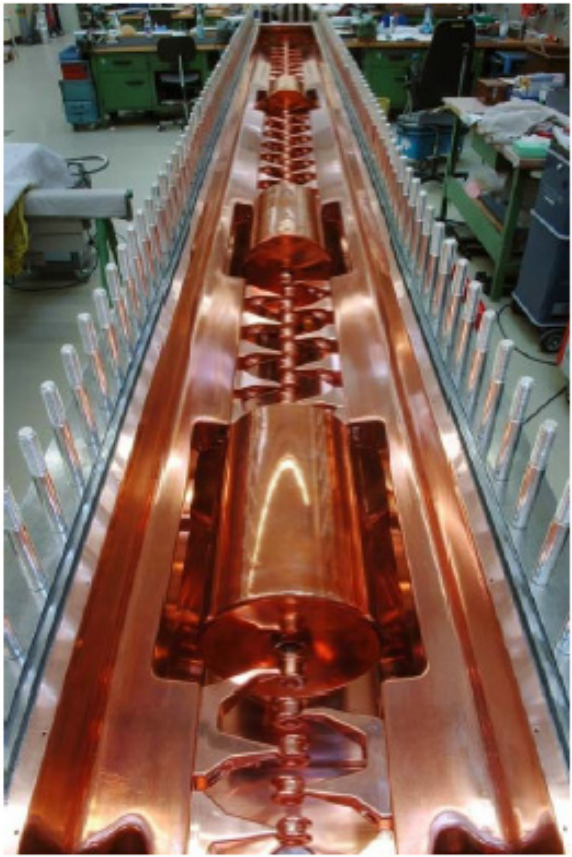
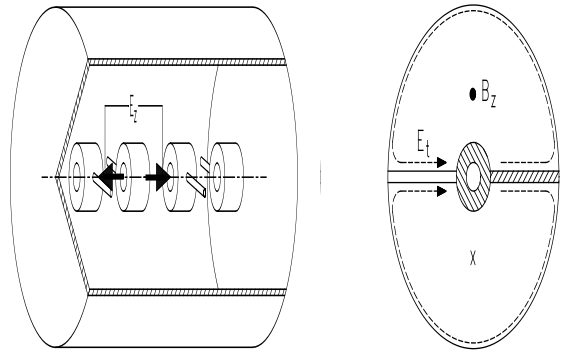
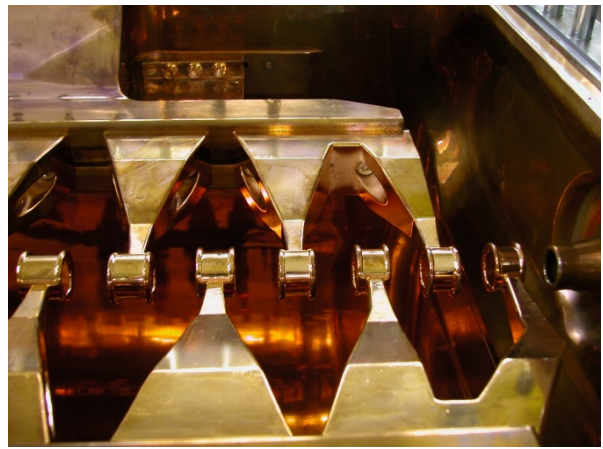

**Fig. 7:** The IH structure

The resonating mode of the cavity is a dipole mode, the TE$_{110}$. The cavity is equipped with thin drift tubes, and alternating the stems on each side of the drift tubes produces a field in the direction of propagation of the beam which accelerates the beam. The focusing is provided by quadrupole triplets located inside the tank in a dedicated section. The IH structure is very efficient in the low-beta region ($\beta$ = 0.02 to 0.08) and at low frequency (up to 200 MHz). It is adapted for low-beta ion acceleration, ideal in a facility which accelerates both protons and carbon ions.

## 2.4 The Drift Tube Linac

The DTL or Alvarez Linac is a cavity resonating in the TM$_{010}$ mode equipped with drift tubes, typically housing quadrupole lenses. A sketch is shown in Fig. 8. In Fig. 9, the field of the fundamental accelerating mode is shown. The length of each drift tube is adapted to the velocity of the beam as the particle has to spend half the RF period inside the drift and half inside the accelerating gap.

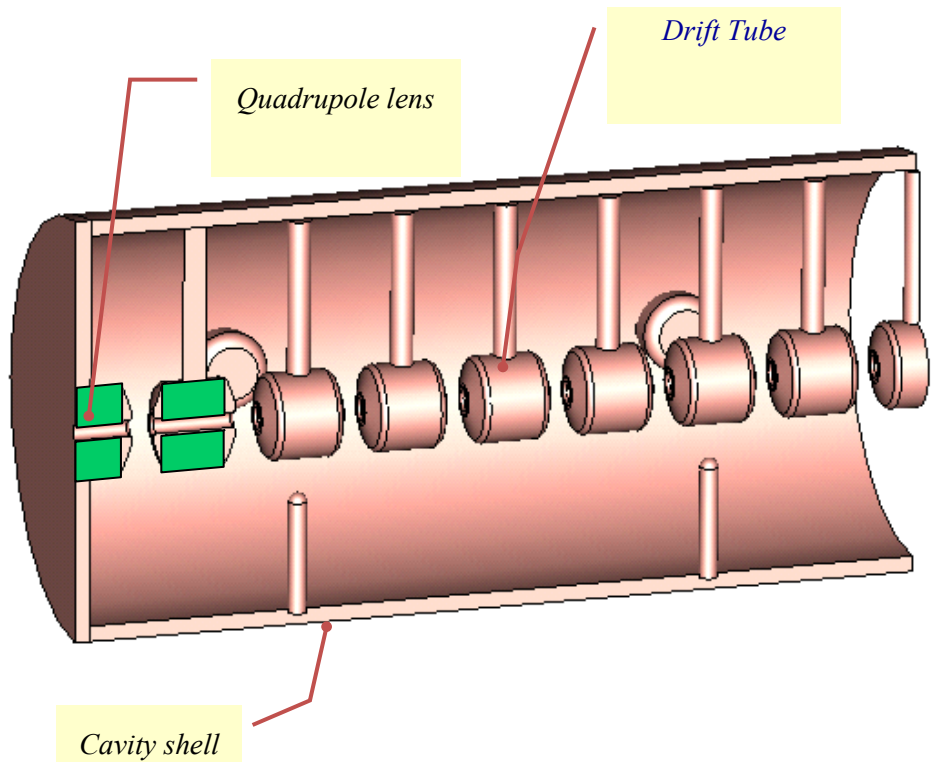

**Fig. 8:** Sketch of the DTL

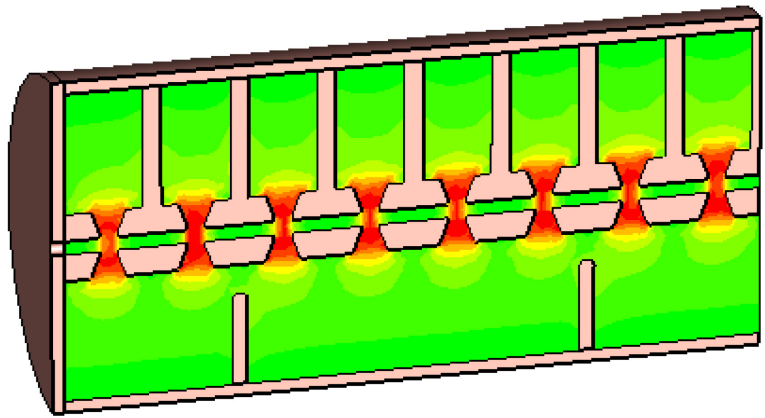

**Fig. 9:** Field distribution in a DTL

## 2.5 The Side-Coupled Linac

Side-Coupled or Cell-Coupled Linacs are strings of cavities operating in the in π/2 mode. Focusing elements are placed outside the cavities and they are completely independent. They are used when the beam is sufficiently energetic.

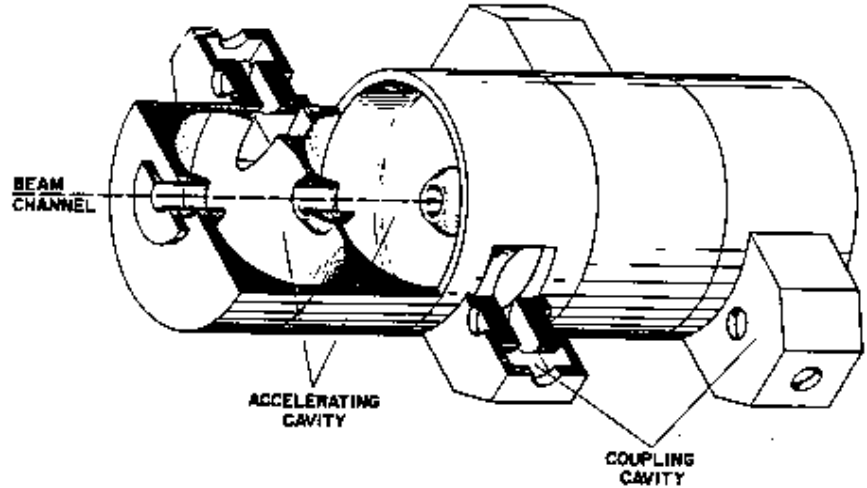

**Fig. 10:** Sketch of SCL

## 3 Quality factors

Each type of structure is adapted for a different energy range and for a different use. In the following, we describe a selection of quality factors to be considered when choosing a structure. A simple sketch of a cavity for illustration purpose s is shown in Fig. 11.

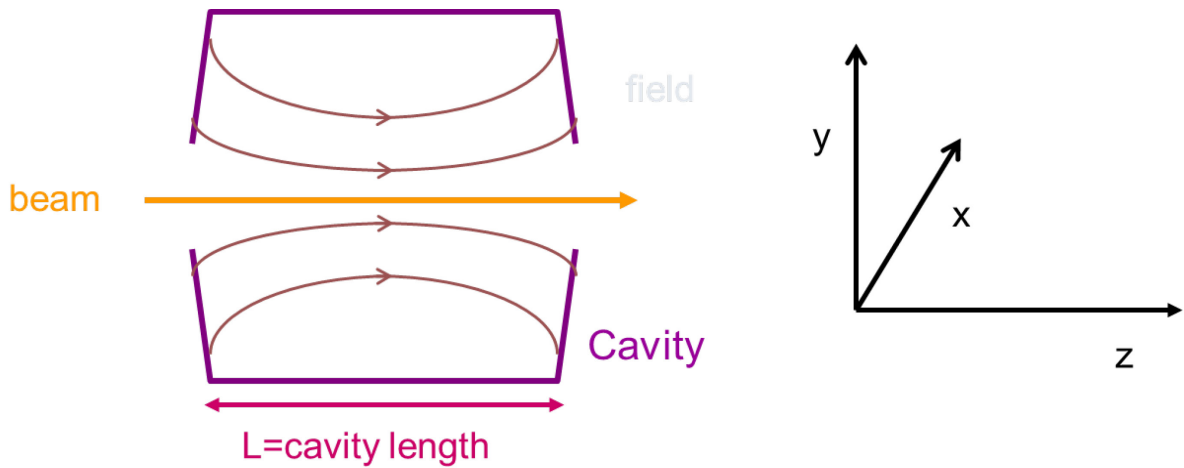

**Fig. 11:** Sketch of an RF cavity

### 3.1 Maximum field/average field

The average electric field (usually indicated with $E_0$ and measured in V/m) is the space average of the electric field along the direction of propagation of the beam in a given moment in time when the field is maximum.

If we write the electric field as

$$E(x,y,z,t) = E(x,y,z)e^{-j\omega t}, \tag{5}$$

and we take the space average

$$E_0 = \frac{1}{L}\int_0^L E_z(x=0, y=0, z)\,\mathrm{d}z,\qquad(6)$$

we obtain a value that gives a measure of how much field is available for acceleration, which depends on the cavity shape, the resonating mode, and the frequency.

The limit to the field in a normal conducting cavity comes from sparking and a useful criterion was determined by W.D. Kilpatrick in 1950 [9], relating the maximum peak surface field to the frequency of a cavity, showing a dependence as in Fig. 12. The maximum surface peak field obtainable depends also on the surface quality and, nowadays, fields up to twice the Kilpatrick limit are obtained. Nevertheless, for a conservative design, a maximum surface field not exceeding 1.7 times the Kilpatrick field is generally advised.

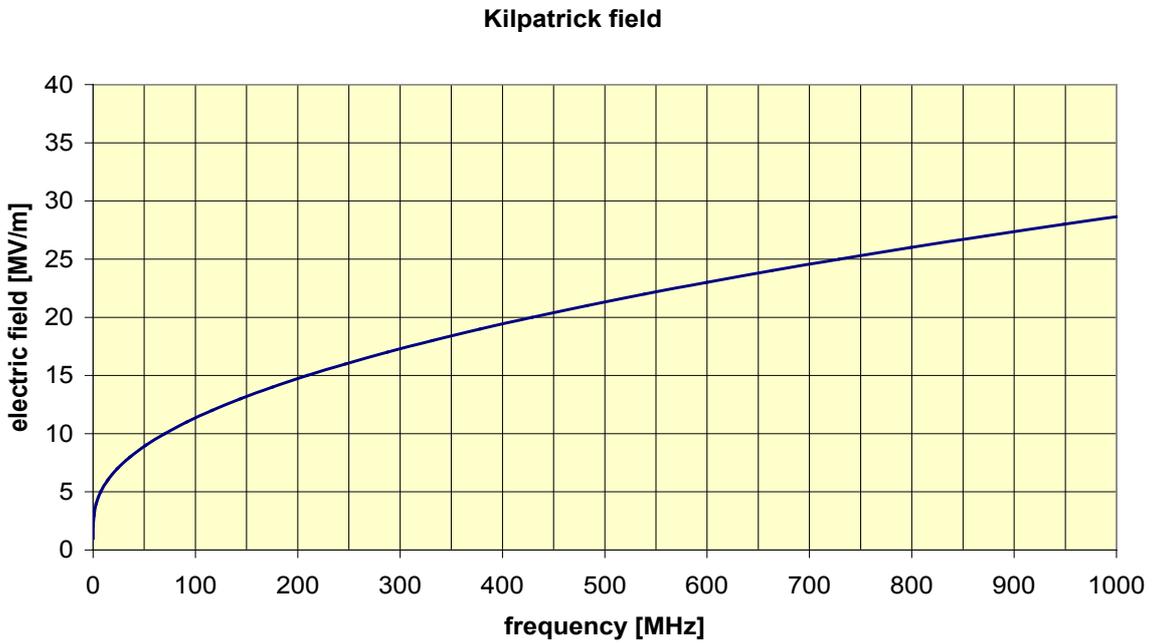

**Fig. 12:** The Kilpatrick criterion

## 3.2 Transit time factor

The transit time factor (indicated with $T$ and dimensionless) is defined as the maximum energy gain per charge of a particle traversing a cavity over the average voltage of the cavity.

To better understand the meaning of this important parameter, let us write the field as

$$E_z(x,y,z,t) = E_z(x,y,z)\,\mathrm{e}^{-\mathrm{i}(\omega t)}\qquad(7)$$

The energy gain of a particle entering the cavity on axis at phase $\varphi$ is

$$\Delta W = \int_0^L q E_z(o,o,z)\,\mathrm{e}^{-\mathrm{i}(\omega t+\varphi)}\,\mathrm{d}z.\qquad(8)$$

Assuming constant velocity through the cavity, we can relate position and time via

$$z = v\cdot t = \beta c t.\qquad(9)$$

We can write the energy gain as
$$\Delta W = qE_0 LT \cos(\varphi) \tag{10}$$

and define the transit time factor as
$$T = \frac{\left|\int_0^L E_z(z) e^{-j\left(\frac{\omega z}{\beta c}\right)} dz\right|}{\int_0^L E_z(z) dz}. \tag{11}$$

*T* depends on the particle velocity and on the gap length. It does not depend on the absolute value of the field.

For a cylindrical pillbox resonating in the TM$_{010}$ mode, assuming a square-wave field distribution, the transit time factor turns out to be
$$T = \frac{\sin\left(\frac{\pi L}{\beta \lambda}\right)}{\left(\frac{\pi L}{\beta \lambda}\right)}. \tag{12}$$

This simple expression tells us that the length of the cavity must be adapted to the energy of the particle to be accelerated, otherwise the efficiency of acceleration can drop drastically.

### 3.3 Effective shunt impedance

The effective shunt impedance (indicated with ZTT and measured in Ω/m) is defined as the ratio of the average effective electric field squared ($E_0 T$) to the power (*P*) per unit length (*L*) dissipated on the wall surface:
$$\text{ZTT} = \frac{(E_0 T)^2}{P} \cdot L. \tag{13}$$

It is independent of the field level and cavity length. It depends on the cavity mode and geometry, and on the velocity of the particle to be accelerated. It is a measure of how much energy a charged particle can gain for 1 W of power when travelling over 1 m of structure. More on the shunt impedance can be found in these proceedings [10].

### 3.4 Comparison of structure

A summary table for the structures discussed in this lecture is reported below (Table 2).

Table 2: Comparison of structures discussed in this lecture

| Type of structure | Ideal range of beta | Frequency | Effective gradient | |
|---|---|---|---|---|
| Radio frequency quadrupole | Low–0.05 | 40–400 MHz | 1 MV/m (350 MHz) | Ions/protons |
| Interdigital H structure | 0.02–0.08 | 40–400 MHz | 4.5 MV/m (200 MHz) | Ions/protons |
| Drift Tube Linac | 0.04–0.5 | 100–400 MHz | 3.5 MV/m (350 MHz) | Ions/protons |
| Cell-Coupled Linac also called Side-Coupled Linac | Ideal Beta=1 But as low as beta 0.3 | 800–3000 MHz | 20 MV/m (3000 MHz) | protons |